\documentclass[conference]{IEEEtran}
\usepackage{amsmath}
\usepackage{algorithmic}
\usepackage{url}
\usepackage{graphicx}
\usepackage{subfigure}
\usepackage{eqnarray}
\usepackage{amssymb}
\usepackage{pifont}
\usepackage{CJK}
\usepackage{lineno}
\usepackage{bm}
\usepackage{multirow}
\usepackage{listings}
\usepackage{xcolor}
\usepackage{times}
\usepackage{tabularx}
\usepackage{balance}
\usepackage{caption}
\usepackage{setspace}
\usepackage{colortbl}
\usepackage{tabu}
\usepackage{bbding} 

\usepackage[linesnumbered,boxed,commentsnumbered,ruled]{algorithm2e}

\newcommand{\cmark}{\ding{51}}%
\newcommand{\xmark}{\ding{55}}%

\newcommand{\figref}[1]{Figure \ref{#1}}
\newcommand{\tabref}[1]{Table \ref{#1}}

\newcommand{\algref}[1]{Algorithm \ref{#1}}

\ifCLASSINFOpdf
\else
\fi

\begin{document}
\title{BU-Trace: A Permissionless Mobile System for Privacy-Preserving Intelligent Contact Tracing}

\author{
\IEEEauthorblockN{
Zhe Peng, Jinbin Huang, Haixin Wang, Shihao Wang, Xiaowen Chu, \\
Xinzhi Zhang, Li Chen, Xin Huang, Xiaoyi Fu, Yike Guo, Jianliang Xu\textsuperscript{$^($\Envelope$^)$}
}
\IEEEauthorblockA{Hong Kong Baptist University\\
\{pengzhe, jbhuang, hxwang, shwang, chxw, lichen, xinhuang, xujl\}@comp.hkbu.edu.hk\\
\{xzzhang2, xiaoyifu, yikeguo\}@hkbu.edu.hk}
}

\maketitle

\begin{abstract}
The coronavirus disease 2019 (COVID-19) pandemic has caused an unprecedented health crisis for the global.
Digital contact tracing, as a transmission intervention measure, has shown its effectiveness on pandemic control.
Despite intensive research on digital contact tracing, existing solutions can hardly meet users' requirements on privacy and convenience.
In this paper, we propose $\mathsf{BU}$-$\mathsf{Trace}$, a novel permissionless mobile system for privacy-preserving intelligent contact tracing based on QR code and NFC technologies.
First, a user study is conducted to investigate and quantify the user acceptance of a mobile contact tracing system.
Second, a decentralized system is proposed to enable contact tracing while protecting user privacy.
Third, an intelligent behavior detection algorithm is designed to ease the use of our system.
We implement $\mathsf{BU}$-$\mathsf{Trace}$ and conduct extensive experiments in several real-world scenarios.
The experimental results show that $\mathsf{BU}$-$\mathsf{Trace}$ achieves a privacy-preserving and intelligent mobile system for contact tracing without requesting location or other privacy-related permissions.

\emph{Keywords}---Privacy-preserving, Permissionless, Intelligent, Contact tracing.

\end{abstract}

\IEEEpeerreviewmaketitle

\section{Introduction}
The eruption of the COVID-19 pandemic has drastically reconstructed the normality across the globe. 
Tracing and quarantine of close contacts is an important and effective non-pharmaceutical intervention (NPI) for reducing the transmission of COVID-19~\cite{flaxman2020estimating}.
A recent study shows that the COVID-19 pandemic can be stopped if 60$\%$ of the close contacts can be  immediately identified, in particular combined with other preventive measures such as social distancing~\cite{ferretti2020quantifying}.
However, traditional manual contact tracing is inefficient because it is limited by a person's ability to recall all the close contacts since infection and the time it takes to reach these contacts.
Thus, in a world coexisting with the infectious coronavirus, an effective and secure digital contact tracing system is much desired.
With such a system, identified close contacts could be provided with early quarantine, diagnosis, and treatment to track and curb the spread of the virus.

There are several challenges in building an effective digital contact tracing system.
First, user privacy protection, especially during contact data collection and processing, is known to have a significant impact on the uptake of such a system~\cite{zeinalipour2020covid, li2012nearby}.
Thus, assuring user privacy should be the very first requirement for digital contact tracing.
Most existing contact tracing systems~\cite{healthCode,koreanApp,indiaApp,safeEntry,bay2020bluetrace} are designed with a centralized model, where personal data is uploaded to a central server for contact matching.
However, in such systems, user privacy could be compromised and system security is not guaranteed. 
Based on the collected personal data (e.g., identities, locations, etc.), the server might be able to infer knowledge pertaining to users' interests.
Moreover, the server is a valuable target for malicious attackers, which may result in serious data breach and leakage.

Second, to fully eliminate concerns about user privacy, contact tracing should be conducted without collecting user location data during normal operations. Even better, the contact tracing mobile app should be \emph{permissionless}, i.e., neither accessing any restricted data nor performing any restricted action that requires users' location permissions. However, most prior solutions either need to access location data such as GPS and e-payment transaction records~\cite{healthCode,koreanApp,indiaApp} or leverage Bluetooth for decentralized contract tracing~\cite{covidSafe,canadaApp,germanApp,swissApp,ens}. As Bluetooth can be used to gather information about the location of a user, the use of Bluetooth must explicitly request the location permission on both Android and iOS platforms, which could affect users' willingness to install the mobile app and participate in contact tracing.

Third, making the contact tracing system \emph{intelligent} is crucial to enhance user experience.
For example, when a user leaves a venue, a check-out reminder would be automatically displayed on the screen to remind the recording of the leaving event.
Moreover, to preserve data privacy and reduce privacy worries, the intelligent algorithm should be localized and realized merely based on permissionless data obtained from the local mobile phone. In other words, advanced sensor data such as step counts cannot be utilized because access to them requires privacy-related permissions, namely the \emph{Activity Recognition} permission on Android and the \emph{Health Data} or \emph{Motion Data} permission on iOS.

\begin{table*}[t]
  \caption{\label{tab:appCompare} Comparison with Global Contact Tracing Apps.}
  \begin{center}
    \begin{tabular}{ |p{3.8cm}<{\centering}|p{2cm}<{\centering}|p{2.4cm}<{\centering}|p{2.1cm}<{\centering}|p{3.0cm}<{\centering}|p{2.1cm}<{\centering}| }
            \hline
            \textbf{Country/Region} & \textbf{Approach} & \textbf{Tracing Technology} & \textbf{Privacy Protection} & \textbf{No Location Permission Needed} & \textbf{System Intelligence} \\
            \hline
            Mainland China (Health Code System) & Centralized & QR Code, GSM, e-Payment Transactions & \xmark & \xmark & \xmark \\
            \hline
            South Korea (Contact Tracing System) & Centralized & GPS & \xmark & \xmark & \xmark \\
            \hline
            India (Aarogya Setu) & Centralized & GPS, Bluetooth & \xmark & \xmark & \xmark \\
            \hline
            Singapore (SafeEntry) & Centralized & QR Code & \xmark & \cmark & \xmark \\
            \hline
            Singapore (TraceTogether) & Centralized & Bluetooth & \xmark & \xmark & \xmark \\

            \hline
            Australia (CovidSafe) & Decentralized & Bluetooth & \cmark & \xmark & \xmark \\
            Canada (COVID Shield) & & & & & \\
            German (Corona-Warn) & & & & & \\
            Switzerland (SwissCovid) & & & & & \\
            Google and Apple (ENS) & & & & & \\
            \hline
            \textbf{BU-Trace} & \textbf{Decentralized} & \textbf{QR Code, NFC} & \cmark & \cmark & \cmark \\
            \hline
    \end{tabular}
  \end{center}
  \vspace{-1em}
\end{table*}

\textbf{Contributions.}
To meet these challenges, we propose $\mathsf{BU}$-$\mathsf{Trace}$, a permissionless mobile contact tracing system based on QR code and Near-Field Communication (NFC) technologies, which simultaneously offers user privacy protection and system intelligence.
Concretely, we make the following major contributions in this paper.

\begin{itemize}

\item  We conduct a user study to investigate and quantify the user acceptance of a mobile contact tracing system.
Specifically, anonymous participants are provided with qualitative virtual focus group interviews and quantitative surveys.
Based on the tailored investigation, we find two most desired properties of the system, \emph{i.e.,} privacy and convenience.
With these user study results, we are inspired and motivated to develop a privacy-preserving intelligent contact tracing system.

\item We propose a decentralized system to enable contact tracing while protecting user privacy.
Compared with other systems, $\mathsf{BU}$-$\mathsf{Trace}$ leverages QR code and NFC technologies to record  users' venue check-in events, which requires no location access permission at all.
Moreover, the system enables participants to confidentially conduct contact matching on local mobile phones based on historical venue check-in records, \emph{i.e.}, no local data is ever exposed during the record matching phase.

\item We design an intelligent behavior detection algorithm to ease the use of the system.
Our algorithm is
$\mathsf{(i)}$ \emph{automatic}, automatically recognizing the user's movement behavior to  display a check-out reminder for recording the leaving event;
and $\mathsf{(ii)}$ \emph{localized}, leveraging the mobile phone's internal accelerator data for local behavior detection.

\item We demonstrate the practicability of $\mathsf{BU}$-$\mathsf{Trace}$ by fully implementing a system using readily-available infrastructural primitives.
We also evaluate the effectiveness of our proposed intelligent behavior detection algorithm within two practical scenarios and hundreds of collected data records.
The experimental results show that $\mathsf{BU}$-$\mathsf{Trace}$ achieves a privacy-preserving and intelligent mobile system for contact tracing without requesting location or other privacy-related permissions.

\end{itemize}

\section{Related Work}

\subsection{Contact Tracing}
Existing approaches for contact tracing utilize various technologies, such as GPS, GSM, Bluetooth, and QR code, to decide a person's absolute and relative location with others.
The current contact tracing systems can be classified into \emph{centralized} and \emph{decentralized} models based on their architectures.
In centralized systems, users are required to upload their data to a central server, which is usually supported by the government authority \cite{healthCode,koreanApp,indiaApp,safeEntry,bay2020bluetrace}.
The server will maintain the user data and perform data matching.
However, user privacy cannot be guaranteed since the user data will be collected and uploaded to the central server.
Oppositely, in decentralized systems, all data will be collected and stored in users' local devices.
Existing decentralized systems mainly utilize the Bluetooth technology to determine the relative distances among users \cite{covidSafe,canadaApp,germanApp,swissApp,ens}.
However, all these approaches need to request location access permission from the mobile operation system.
Concerning user privacy, users may not be willing to install and use these mobile apps, which makes the contact tracing ineffective. For example, TraceTogether~\cite{bay2020bluetrace}, the Bluetooth-based contact tracing app from Singapore, has only 1.4M active users (25\% of population) after more than five months of release~\cite{cna2020news}.
For a more comprehensive comparison, we summarize and contrast the features of various contact tracing systems in \tabref{tab:appCompare}.
Different from existing systems, our proposed $\mathsf{BU}$-$\mathsf{Trace}$ in this paper leverages QR code and NFC technologies, which are permissionless.
Moreover, the system simultaneously enables user privacy protection and system intelligence.

\subsection{Sensor Data Analysis}
Many new types of sensors have been equipped with modern mobile phones, which can produce abundant sensing data about environment and human.
In order to effectively analyze the sensor data to obtain valuable information, different methods have been proposed \cite{peng2018indoor,gonzalez2017direct,yao2017efficient}.
Specifically, in many human-centric intelligent applications, it is very significant to accurately recognize various user behaviors and motions from the sensor data.
Among existing techniques, machine learning is proven to be a promising solution for processing time sequence data \cite{shi2018machine}.
Many learning-based methods, such as recurrent neural networks (RNN) and long short term memory (LSTM), have been developed and applied in real-world scenarios successfully \cite{medsker2001recurrent,hochreiter1997long}.
Santos et al. proposed a method to model human motions from inertial sensor data \cite{santos2019artificial}.
Shoaib et al. implemented a smartphone-based data fusion technique for detecting various physical activities \cite{shoaib2014fusion}.
Bedogni et al. came up with a solution to detect users' motion types in realtime by using sensor data collected from smartphones \cite{bedogni2012train}.
In addition, some studies have attempted to leverage sensor data to detect users' transportation modes.
Fang et al. proposed a new method to classify different transportation modes by analyzing sensor data from multiple sources \cite{fang2016transportation}.
In order to effectively recognize various human activities, most existing approaches need to collect abundant sensor data and conduct computation on a powerful server.
In contrast, considering user privacy, our approach will conduct sensor data analysis on local mobile phones.

\section{User Study}
With the voluntary principle, in order for $\mathsf{BU}$-$\mathsf{Trace}$ to be widely adopted in communities to break the virus transmission chain, the developed approach should be able to effectively satisfy users' demands and concerns.
Therefore, we start by understanding users' perception and acceptance of a mobile contact tracing system.

\begin{table}[t]
\caption{For positive participants (n = 13), the reasons of stopping using the app (1 = the most disagree; 5 = the most agree).}
\centering
\begin{tabular}{p{3cm}|c|c|c|c|c|c|c}
\hline
                                      & Mean          & SD   & 1 & 2 & 3 & 4 & 5 \\
\hline
If privacy leakage is found           & \textbf{4.23} & 1.09 & 0 & 1 & 3 & 1 & 8 \\
If the app is not convenient to use   & \textbf{3.85} & 0.80 & 0 & 0 & 5 & 5 & 3 \\
\hline
\end{tabular}
\label{tab:StudyPositiveNotUse}
\vspace{-1em}
\end{table}

\subsection{Method}
A user study involving quantitative survey and qualitative focus group interviews were carried out in September 2020.
We recruited 20 participants (8 male, 12 female, aged 18-55 years) via the purposive sampling method.
Invitation letters for joining the study were sent to the Deans and Department Heads within a public university in Hong Kong, asking the Deans and Department Head to nominate three representatives from the Faculty.
The participants included 8 university students and 12 staffs from five different Schools and Faculties, covering a wide range of disciplines (including science, social science, arts, and humanities).

The user study was conducted online.
The participants received a link directed to the quantitative questionnaire.
After finishing the questionnaire, they were asked to join a virtual focus group hosted on Slack.
The participants were allocated into four different discussion channels (each channel included four to five participants) and could share their perceptions and concerns on a mobile contact tracing system, and discuss with other participants in an asynchronous manner.

The user study mainly focused on examining the factors that may make people adopt or resist the mobile contact tracing application.
A total of 18 filter questions were settled, which meant different questions were displayed to the participants based on their answer of the current question.
For all the quantitative questions, five-point Likert scales were used where 1 meant the most disagreement, and 5 meant the most agreement.

\subsection{Results}

To comprehensively investigate the user acceptance of a mobile contact tracing system, participants were divided into two groups based on whether they hold a positive or negative view towards the system.
Concretely, 13 out of 20 participants (called \emph{positive participants}) indicated that they would like to install and use such a mobile app for getting virus risk alerts, while others (called \emph{negative participants}) indicated that they would not.

For positive participants, they were first requested to input the most important factors that make them use such a mobile app.
Among their answers, the factors frequently mentioned include getting alerts, receiving latest news about the pandemic, and self-protection.
In addition, participants were asked to input factors that might make them stop using the app.
Privacy concern and inconvenience were two mentioned factors.
Furthermore, quantitative questions were provided to evaluate the reasons of stopping using the app.
As shown in \tabref{tab:StudyPositiveNotUse}, both privacy and convenience are crucial factors (mean values are larger than 3).
Based on the survey for positive participants, we find that interviewees hope to receive timely infection risk alerts from the app, but they have certain privacy and inconvenience concerns.

\begin{table}[t]
\caption{For negative participants (n = 7) subjects, the reasons of changing the idea of the app usage (1 = the most disagree; 5 = the most agree).}
\centering
\begin{tabular}{p{3cm}|c|c|c|c|c|c|c}
\hline
                                     & Mean          & SD   & 1 & 2 & 3 & 4 & 5 \\
\hline
If privacy is completely protected   & \textbf{3.14} & 0.38 & 0 & 0 & 6 & 1 & 0 \\
If the app is convenient to use      & \textbf{3.29} & 0.76 & 0 & 1 & 3 & 3 & 0 \\
\hline
\end{tabular}
\label{tab:StudyNegativeUse}
\vspace{-1em}
\end{table}

\begin{table*}[t]
\caption{Evaluation on options of permitting the app to check in and check out a venue, for all the participants (n = 20) (1 = the most disagree; 5 = the most agree).}
\centering
\begin{tabular}{cl|c|c|c|c|c|c|c}
\hline
                         & \multicolumn{1}{c|}{} & Mean & SD   & 1 & 2 & 3 & 4 & 5 \\
\hline
\multicolumn{1}{c|}{\multirow{2}{*}{Check In}} & Scan the QR code every time (no location access)   & \textbf{3.55} & 1.15 & 1 & 3 & 4 & 8 & 4 \\
\multicolumn{1}{c|}{}    & Bluetooth for auto check-in (need location access)      & \textbf{2.95} & 1.23 & 2 & 6 & 6 & 3 & 3 \\
\hline
\multicolumn{1}{c|}{\multirow{3}{*}{Check Out}} & Scan the QR code every time (no location access)   & \textbf{3.15} & 1.23 & 2 & 4 & 6 & 5 & 3 \\
\multicolumn{1}{c|}{}    & Bluetooth for auto check-in (need location access)      & \textbf{3.05} & 1.32 & 3 & 4 & 5 & 5 & 3 \\
\multicolumn{1}{c|}{}    & AI for auto check-out (no location access)      & \textbf{3.40} & 1.39 & 3 & 2 & 4 & 6 & 5 \\
\hline
\end{tabular}
\label{tab:StudyAI}
\vspace{-1em}
\end{table*}

For negative participants, they were first requested to grade two scenarios where they might change the idea of using such an app.
As shown in \tabref{tab:StudyNegativeUse}, privacy concern and inconvenience related conditions both received higher grades.
After that, participants were invited to input some suggestions to increase the adoption rate of the app.
Participants express very strong and salient opinions on the privacy issue and data usage.
Some incentive mechanisms were also proposed, such as free drink or food coupons, free face masks, and free COVID-19 testing once getting a virus risk alert.
Based on this survey, we find that privacy concern and inconvenience are also important concerns for negative interviewees.

Finally, for all participants, we provided some quantitative questions about options of permitting the app to check in and check out a venue, e.g., the canteen.
For the check-in scenario, users can 1) scan a QR code every time when entering a venue (this option does not require the permission to access the user's location) or 2) use Bluetooth for auto check-in (this option requires authorizing the location permission, though the app does not collect the user's location data).
As shown in \tabref{tab:StudyAI}, users preferred to scan a QR code every time instead of using Bluetooth for auto check-in.
For the check-out scenario, besides the QR code and Bluetooth options, another option was provided, i.e., using intelligent technologies for auto check-out (this option does not require the permission to access the user's location).
As shown in \tabref{tab:StudyAI}, participants preferred to choose intelligent technologies to estimate the duration of the stay in a particular venue and realise the auto check-out.

Therefore, based on the tailored investigation, we find two most desired properties of a contact tracing system, i.e., privacy and convenience.
With these user study results, we are inspired and motivated to develop an effective contact tracing system.

\section{System Overview}
\begin{figure}[t]
    \centering
    \includegraphics[width=0.45\textwidth]{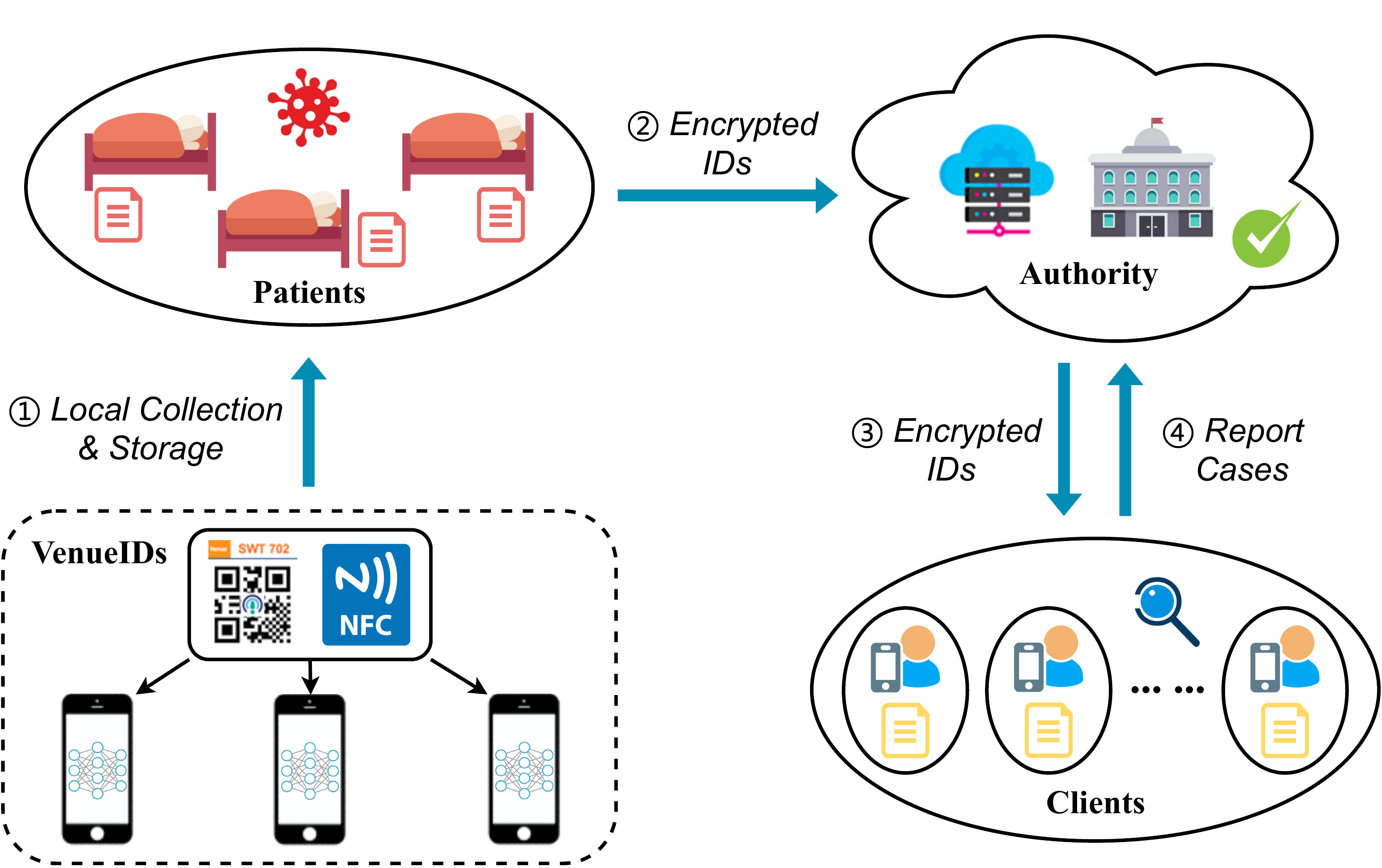}
    \caption{The system architecture of $\mathsf{BU}$-$\mathsf{Trace}$}
    \label{fig:system}
    \vspace{-1em}
\end{figure}

$\mathsf{BU}$-$\mathsf{Trace}$ is a permissionless mobile system for privacy-preserving intelligent contact tracing. The system will send an alert message to users through a mobile app if they and an infected person have visited the same place within a time period that gives rise to risks of exposure.
\figref{fig:system} shows the $\mathsf{BU}$-$\mathsf{Trace}$ system architecture.
Our system mainly consists of the following actors: $\mathsf{(i)}$ \emph{patient}, $\mathsf{(ii)}$ \emph{client}, and $\mathsf{(iii)}$ \emph{authority}.
A patient is a virus-infected person, while a client refers to an unconfirmed person.
An authority represents a government sector or an organization, which can provide a close proximity certification for further screening.
In the following, we briefly describe three basic modules of the system.

\emph{Permissionless location data collection.}
$\mathsf{BU}$-$\mathsf{Trace}$ utilizes QR code and NFC technologies to record users' venue check-in information. When users scan the system's QR code or NFC tag before entering a venue, the venue ID is collected. Then, the collected venue ID, as well as the time of the visit, are saved on users' mobile phones.
Different from existing systems, our app does not request the location permission from the mobile platforms but only a camera usage notification (no location access) will be displayed on the screen during the first attempt, which safeguards user privacy from the system level.

\emph{Privacy-preserving contact tracing.}
The contact tracing module is designed as a decentralized approach to protect user privacy.
In this module, confirmed patients need to upload their venue visit records within the past 14 days to the authority.
Specifically, to enhance the security, the hashed value of a venue ID instead of the plain text is transmitted to the authority.
The authority will further broadcast the encrypted venue records uploaded by confirmed patients to other clients.
Upon receiving the encrypted venue records, all clients apply the same hash function to their own venue records and conduct a cross-check on local mobile phones.
If a match is found, the app will display an alert notification and the client could report his/her case to the authority for further follow-up.
In the whole process, both data storage and data computation are conducted in a decentralized manner to protect user privacy and make the system more scalable.

\emph{Intelligent behavior detection.}
We design an automatic check-out function based on an intelligent behavior detection module, which improves not only user experience but also time accuracy of check-out records.
To avoid the location permission request and alleviate users' privacy concerns, the intelligent behavior detection module utilizes only inertial sensor data from the mobile phone's accelerometer.
We design a simple yet effective sliding window-based detection algorithm to detect the behavior transition for auto check-out reminders.
Overall, the data analysis procedure is strictly restrained in local mobile phones based on sensor data that does not request users' location or any other privacy-related permissions. We give more details about the intelligent behavior detection method in the next section.

\section{Intelligent Behavior Detection}

For effective contact tracing, both venue check-in and check-out events should be precisely recorded. $\mathsf{BU}$-$\mathsf{Trace}$ enables users to record the check-in event via scanning the QR code or tapping the NFC tag when they enter a venue. However, our pilot experiments showed that users could easily forget to record the check-out event when they left the venue. Therefore, we design an intelligent behavior detection method to facilitate the recording of check-out events.

\begin{figure}[t]
    \centering
    \includegraphics[width=0.45\textwidth]{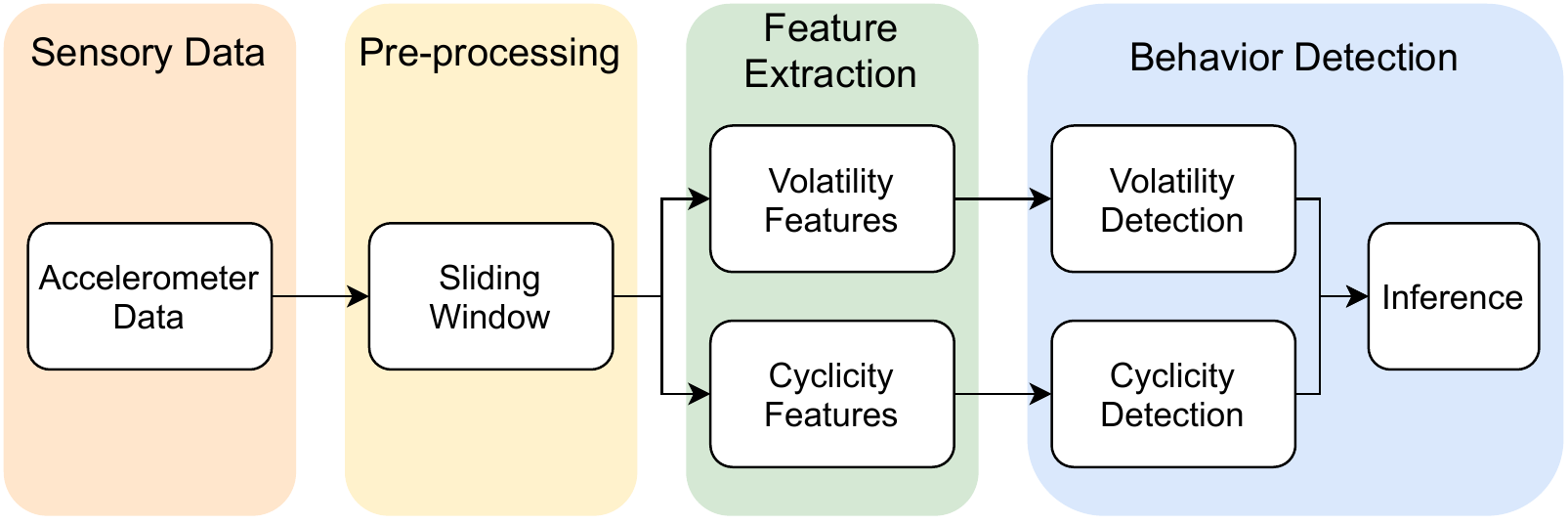}
    \caption{The framework of accelerometer-based intelligent behavior detection}
    \label{fig:behaviorDetection}
    \vspace{-1em}
\end{figure}

\subsection{Accelerometer-Based Behavior Detection Framework}

In this section, we introduce our accelerometer-based intelligent behavior detection framework for auto check-out reminders in $\mathsf{BU}$-$\mathsf{Trace}$, as illustrated in \figref{fig:behaviorDetection}.

\begin{figure}[t]
    \centering
    \includegraphics[width=0.45\textwidth]{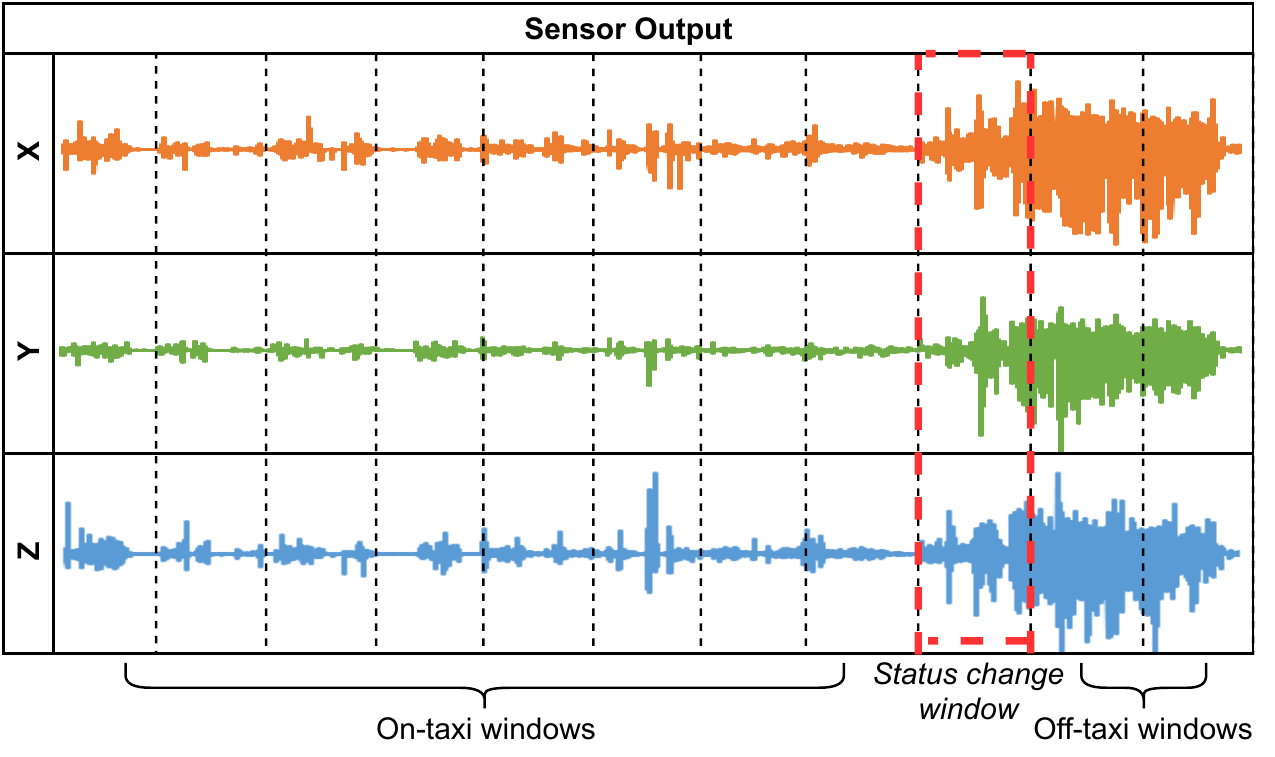}
    \caption{Sensor data pre-processing based on sliding window}
    \label{fig:slidingWindow}
    \vspace{-1em}
\end{figure}

Our hypothesis is that there is usually a behavior transition when a user is leaving a venue (e.g., taxi, train, restaurant, theatre, etc.).
$\mathsf{BU}$-$\mathsf{Trace}$ monitors the accelerometer data in three orthotropic directions (i.e., X, Y, and Z axes, denoted as $\mathbf{A_x}$, $\mathbf{A_y}$, and $\mathbf{A_z}$, respectively) and intelligently infers the check-out event by detecting such a transition.
To reduce the power consumption of $\mathsf{BU}$-$\mathsf{Trace}$, we design a simple yet effective sliding window-based algorithm to detect the behavior transition. As shown in \figref{fig:slidingWindow}, the original sensor data is separated into different windows (denoted by dashed line) with a window width $l$.
For each window, taking the taxi ride as an example, it will be classified into two categories with a behavior detection algorithm, i.e., on-taxi window and off-taxi window.
As such, the inferred check-out time will be located in the first recognized off-taxi window (highlighted by red dashed rectangle).

\subsection{Behavior Detection Algorithm}
\algref{alg:algorithm} describes the procedure of the behavior detection algorithm.
The designed algorithm mainly consists of two behavior detection methods (i.e., volatility detection and cyclicity detection) in terms of $\mathsf{(i)}$ \emph{walking volatility} and $\mathsf{(ii)}$ \emph{walking cyclicity}, respectively.
Then, the final behavior detection result is jointly decided by the two methods.

\textbf{Volatility detection.}
In the volatility detection method, we aim to identify the change of behavior patterns caused by the check-out action.
In many practical scenarios, the check-out action happens in a short period and is followed by continuous walking for a period of time (e.g., getting off a taxi, or leaving a restaurant). These actions will cause a distinct change of sensor readings in the time domain. With this intuition, we propose a simplified polynomial logistic regression algorithm to classify a time window. First, we use powered average to intensify the features (line \textbf{8} in \algref{alg:algorithm}).
For a specific window, the powered average $\mathbf{\Bar{P}}$ of accelerometer readings in the three axes is calculated as
\begin{equation}
\mathbf{\Bar{P}} = \frac{1}{l} \sum\limits_{i=1}^{l} [(A_x^i)^k + (A_y^i)^k + (A_z^i)^k],
\label{equ:poweredAverage}
\end{equation}
where $l$ is the window width, $k$ is an even integer hyperparameter, $A_x^i$ (resp. $A_y^i$, $A_z^i$) is the $i$th accelerometer reading in the window along the X (resp. Y, Z) axis. Specifically, $k$ is constrained to be even in order to generate positive values. Through extensive experiments, we find that $k=4$ is a good choice in practice.
Then, we identify the check-out window if $\mathbf{\Bar{P}} > \mathbf{L_a}$ (line \textbf{12} in \algref{alg:algorithm}), where $\mathbf{L_a}$ is a threshold learnt from our collected training data.

\textbf{Cyclicity detection.}
In the cyclicity detection method, we aim to recognize the unique and inherent behavior pattern of human walking.
Based on our observation, the accelerometer readings of human walking usually present cyclicity. Thus, leveraging this property, we can recognize the walking windows for the check-out event through accurately capturing the periodic crests.
With this intuition, we first strengthen the cyclicity features through the $k$th power of the original accelerometer data.
Then, in order to reduce the influence of unintentional shakes, we further use wavelet transform \cite{rein2010low} to filter the false crests in the frequency domain (line \textbf{10} in \algref{alg:algorithm}).
Specifically, for the accelerometer readings in each window $A_{win}$, we process the $k$th power sequence $A_{win}^k$ through
\begin{equation}
\mathbf{W_A} = \frac{1}{\sqrt{|a|}} \sum\limits_{i} A_{win}^k(i) \overline{\Psi \left( \frac{t-b}{a} \right)},
\label{equ:poweredAverage2}
\end{equation}
where $a = a_0^m$ ($a_0 > 1$, $m \in Z$) is the scale parameter, $b = n \cdot a_0^m \cdot b_0$ ($b_0 > 0$, $n \in Z$) is the translation parameter, $\overline{\Psi \left( t \right)}$ is the analyzing wavelet with a conjugate operation.
Moreover, based on the output $\mathbf{W_A}$, we detect the wave crests and count the amount $\mathbf{P_c}$ with the constraints of peak value $p_v$ and peak interval $p_i$ (line \textbf{11} in \algref{alg:algorithm}).
Finally, a threshold $\mathbf{L_p}$ for the amount of crests, which is learnt from training data, is used to classify the window. If $\mathbf{P_c} > \mathbf{L_p}$, this window will be categorized as the walking window (line \textbf{12} in \algref{alg:algorithm}).

\textbf{Joint detection.}
To maximize the detection accuracy of check-out events, we combine the volatility detection and cyclicity detection methods.
For each window of accelerometer readings, it will be finally inferred as a check-out window, only if both methods indicate the positive.
Thus, in the sliding window framework, a boolean sequence $\mathcal{S}$ will be derived for each accelerometer data record, for example, $\mathcal{S} = \{ \ldots, in, in, in, out, out, out, \ldots \}$.
Our joint detection algorithm then uses a variable $\mathbf{C_w}$ to record the number of continuous check-out windows in the boolean sequence $\mathcal{S}$ (line \textbf{13} in \algref{alg:algorithm}).
Finally, once $\mathbf{C_w} > \mathbf{L_w}$ (line \textbf{16} in \algref{alg:algorithm}), where $\mathbf{L_w}$ is an indicator threshold to infer the check-out event, a check-out reminder will be triggered in the mobile app.

Using the training data collected in the real-world, we employ grid searches to automatically find the optimal values of the hyper-parameters and thresholds used in the behavior detection algorithm.

\begin{algorithm}[t]
    \caption{Sliding Window-based Joint Detection Algorithm}
    \label{alg:algorithm}
    \SetAlgoLined
    \KwIn{Accerelometer readings $A$ containing $A_x$, $A_y$, $A_z$\;
    Sliding window width $l$\;
    Powered average threshold $L_a$\;
    Crest amount threshold $L_p$\;
    Continuous check-out window amount threshold $L_w$\;}
    \KwOut{A boolean value $T$ indicating whether the user has checked out}

    \While{there is input $A$}{
        Window count $C_w = 0$\;
        Empty window $A_{win}=\{\}$\;
        \While{$|A_{win}| <$ $l$ }{
            $A_{win}$.append($A$)\;}
        \For{$A_i$ in $A_{win}$}{
        Calculate Powered Average $\Bar{P}$ += $\frac{(A_x^i)^k + (A_y^i)^k + (A_z^i)^k}{|A_{win}|}$\;
        }
        $W_A \leftarrow{}$ WaveletTransform($A_{win}^k$)\;
        $P_c \leftarrow{}$ FindCrests($W_A$)\;
        \uIf{$\Bar{P} > L_a$ \&\& $P_c > L_p$}{
            $C_w += 1$ \;
        }
        \uElse{
            $C_w = 0$ \;
        }
        \uIf{$C_w > L_w$}{
            return True\;
        }
    }
\end{algorithm}

\begin{figure*}[t]
  \centering
  \begin{tabular}{ccc}
      \multicolumn{3}{c}{\includegraphics[width=0.6\textwidth]{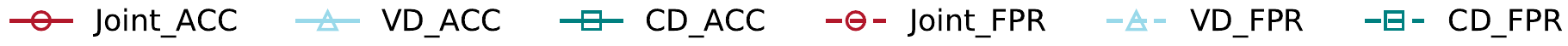}} \\
      \includegraphics[width=0.32\textwidth]{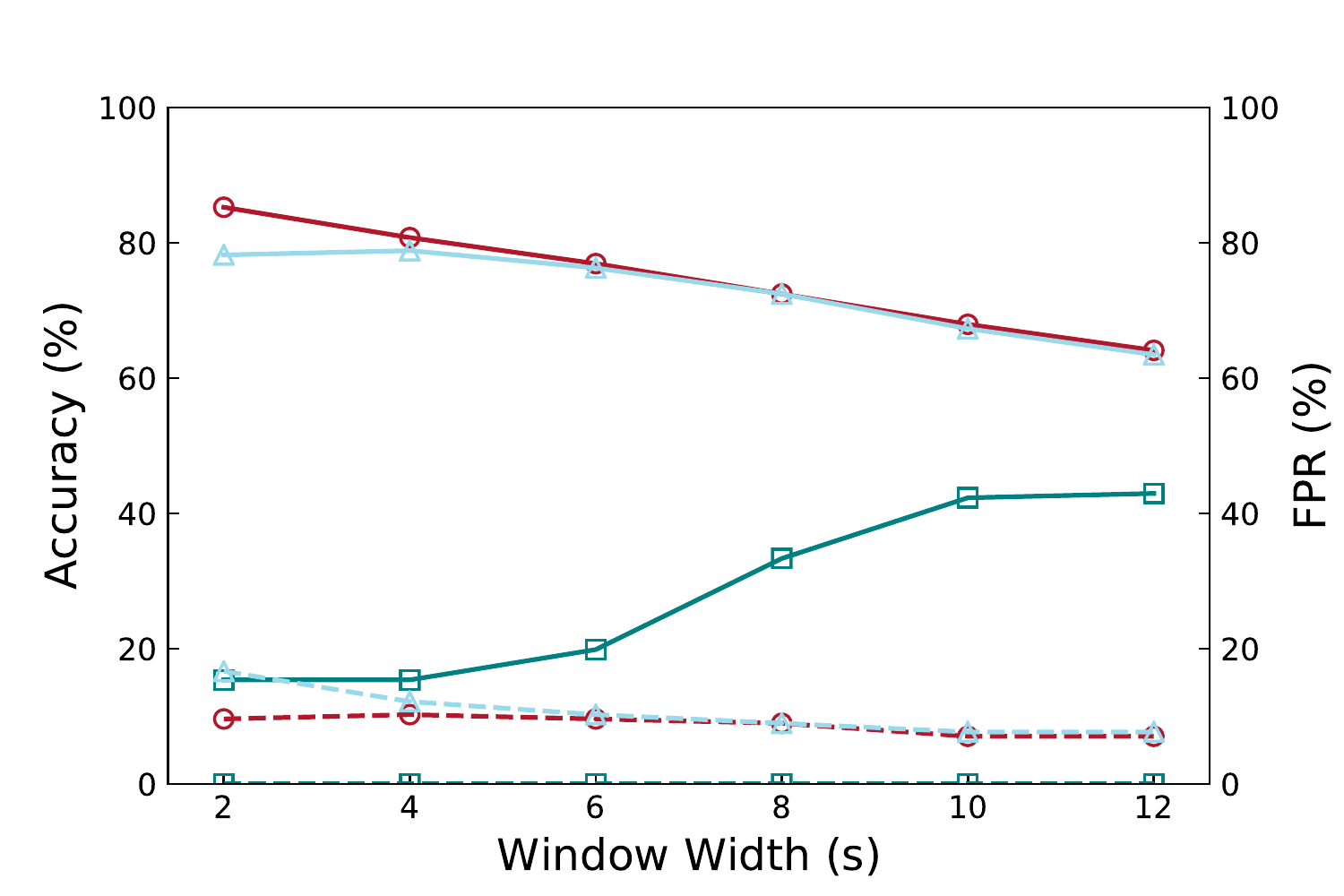} &
      \includegraphics[width=0.32\textwidth]{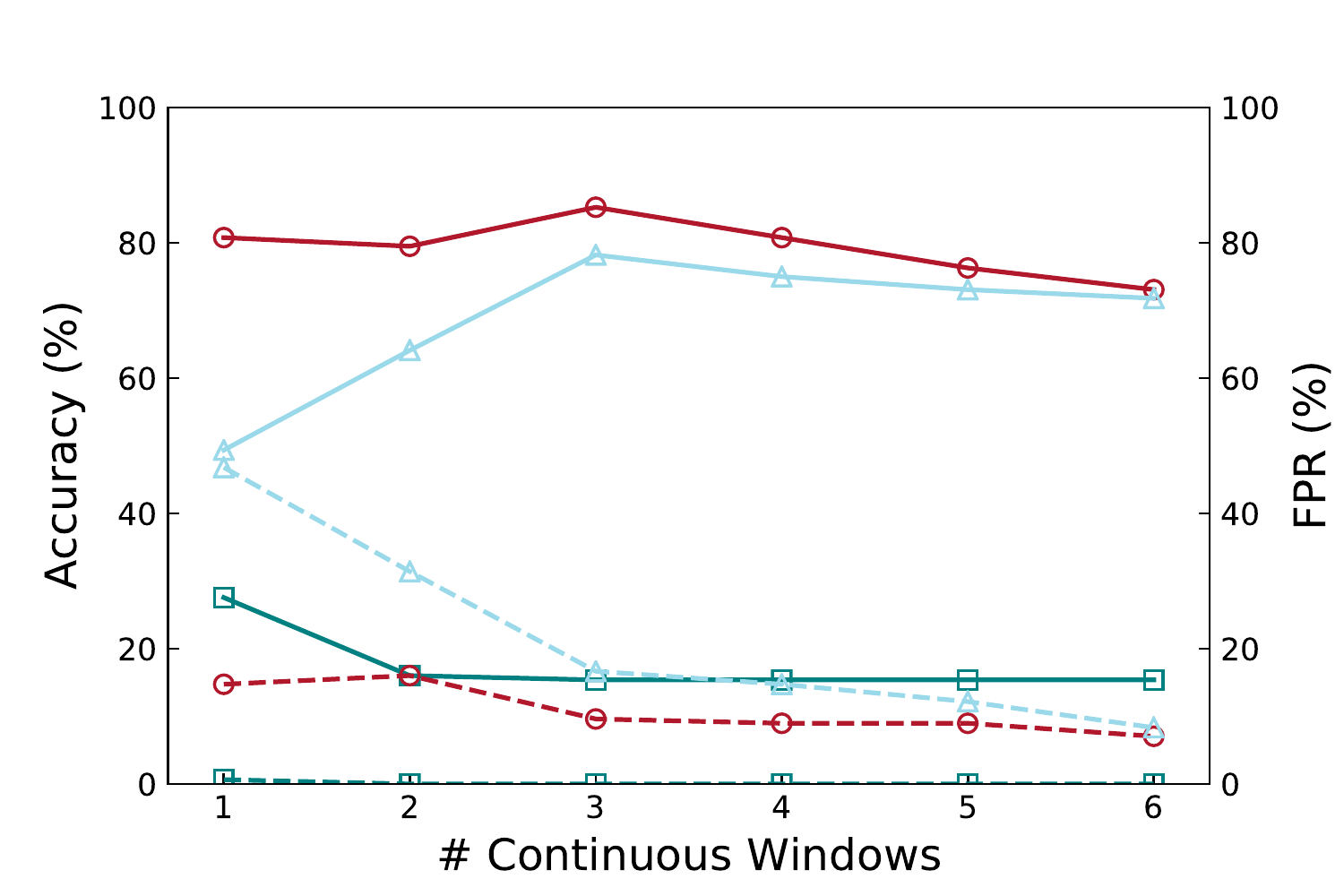} & 
      \includegraphics[width=0.32\textwidth]{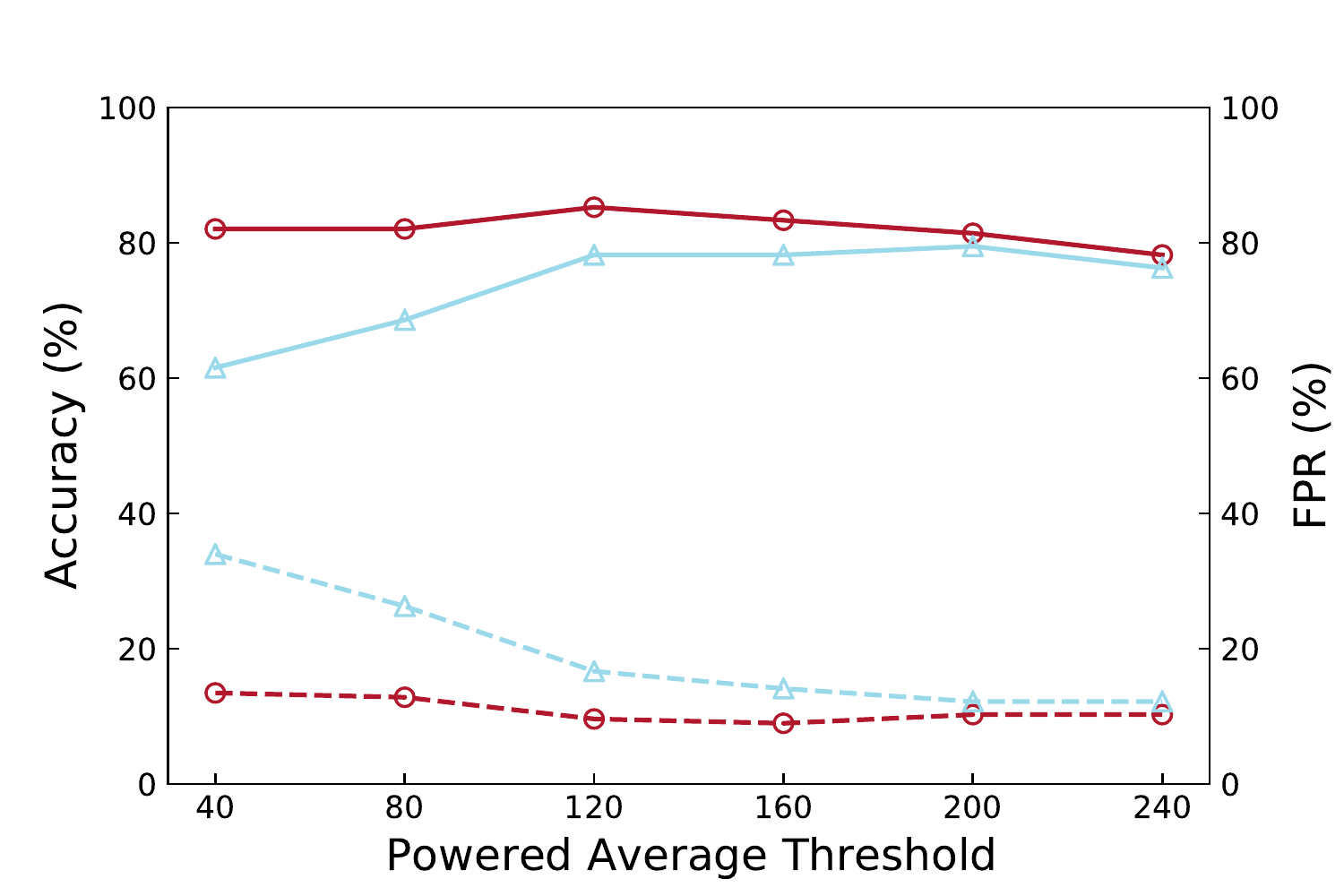} \\
      (a) Effect of $l$ (Taxi). & (b) Effect of $\mathbf{L_w}$ (Taxi). & (c) Effect of $\mathbf{L_a}$ (Taxi). \\
      \includegraphics[width=0.32\textwidth]{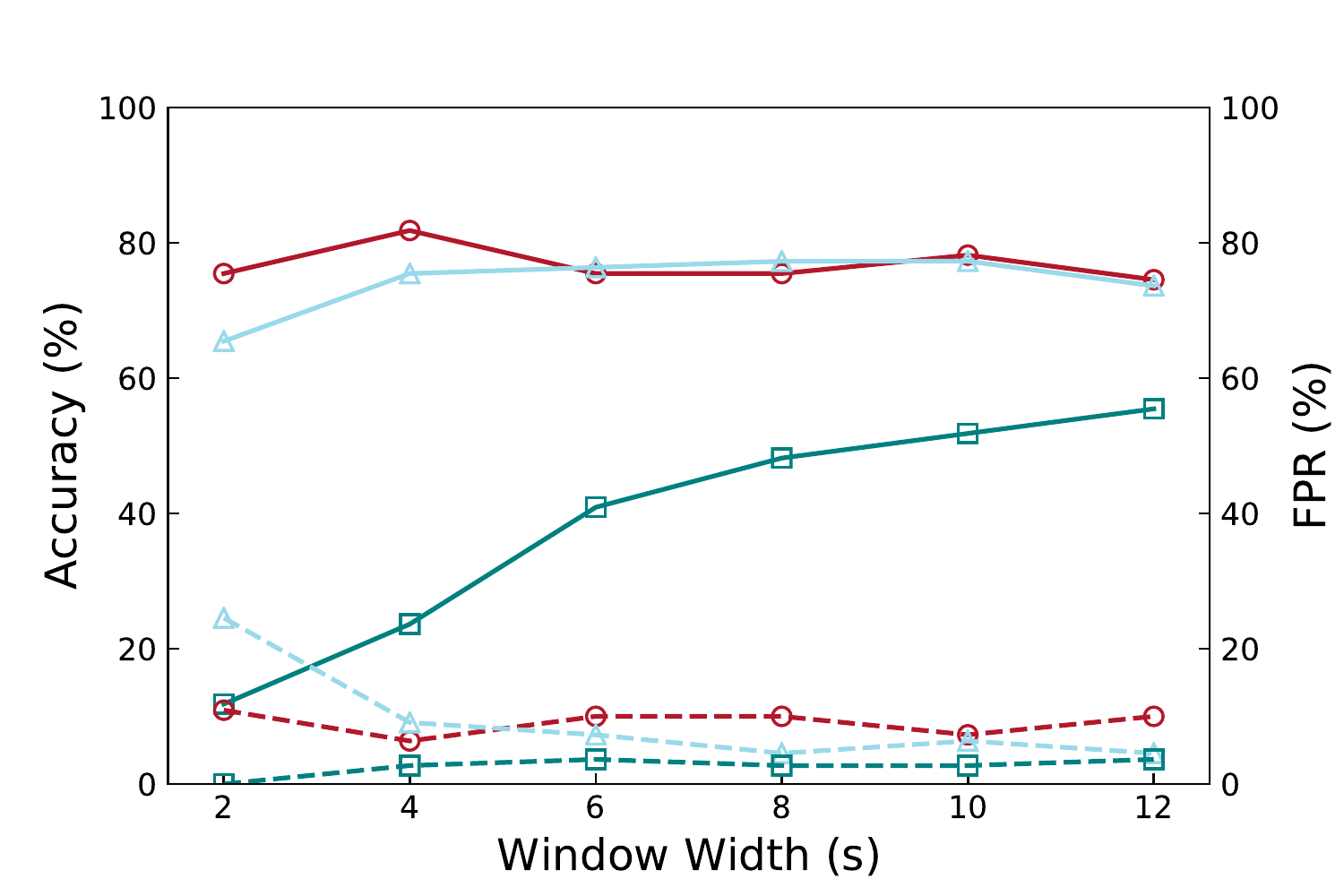} & 
      \includegraphics[width=0.32\textwidth]{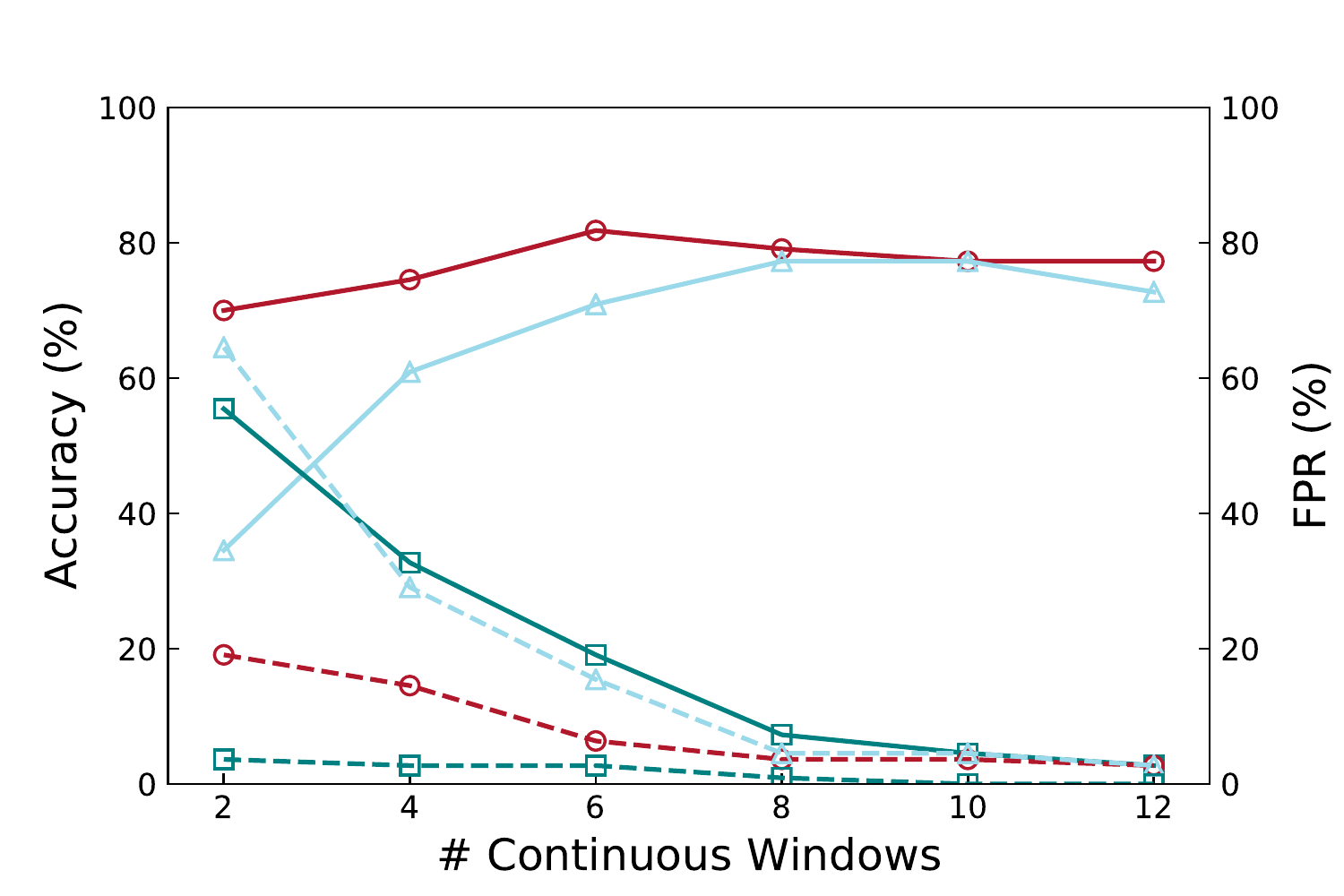} & \includegraphics[width=0.32\textwidth]{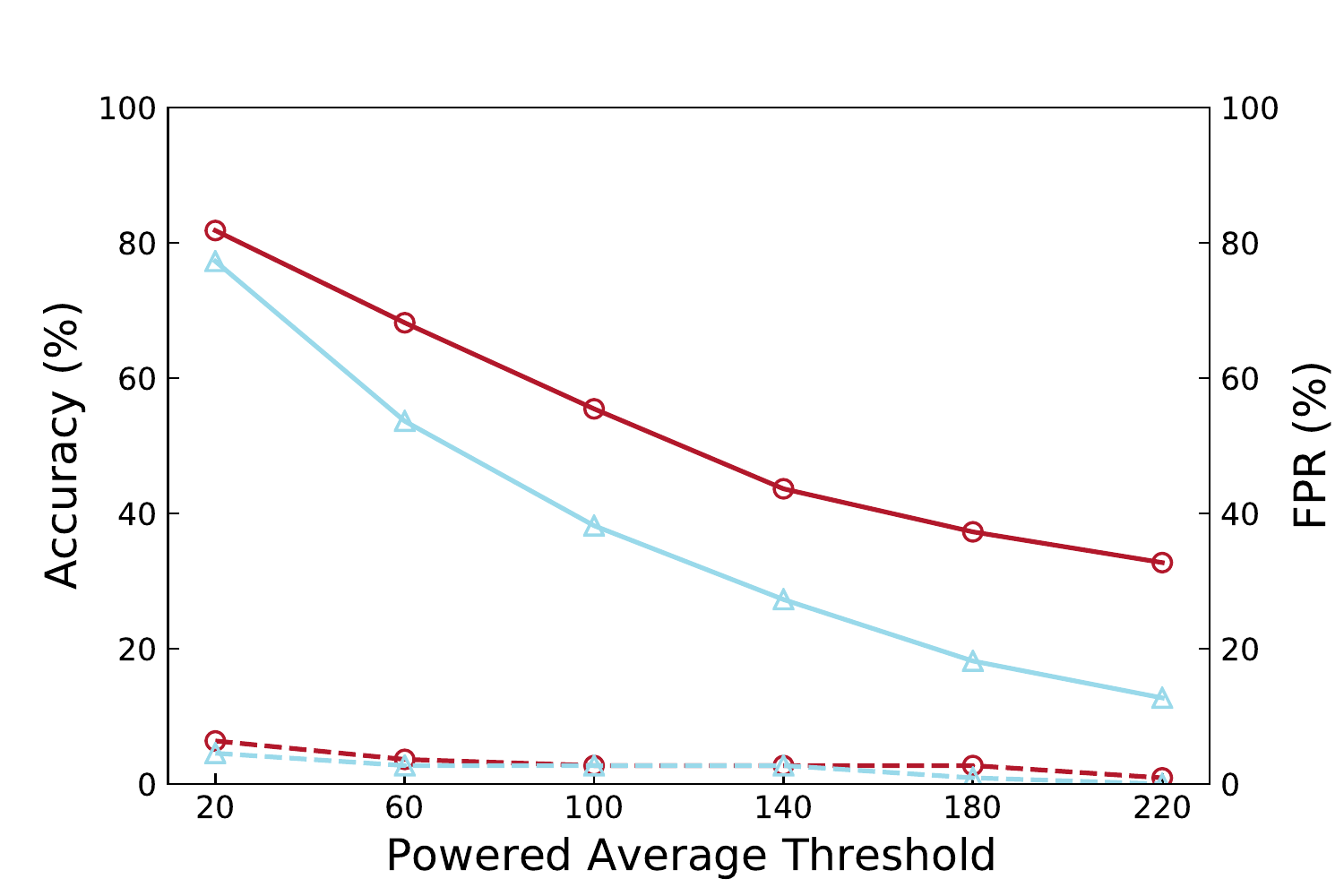} \\
      (d) Effect of $l$ (Canteen). & (e) Effect of $\mathbf{L_w}$ (Canteen). & (f) Effect of $\mathbf{L_a}$ (Canteen). \\
  \end{tabular}
  \caption{Effectiveness evaluation of intelligent behavior detection algorithms.}
  \label{fig:parameters}
  \vspace{-1em}
\end{figure*}

\section{Implementation and Evaluation}

\subsection{Experimental Setup}
We implement the $\mathsf{BU}$-$\mathsf{Trace}$ mobile app with Java and the back-end system with PHP.
Four models of smartphones are used for evaluation, including Samsung A715F, Samsung A2070, Xiaomi 10 Lite 5G, and OPPO Reno4 Pro 5G.
We conduct experiments in two representative real-world scenarios, i.e., taxi and canteen.
Specifically, 156 and 110 accelerometer data records are collected in the taxi scenario and the canteen scenario, respectively.
The data record length ranges from 1 minute 48 seconds to 39 minutes 40 seconds.
The sampling frequency of the accelerometer sensor is set at 50Hz.

To evaluate our proposed intelligent behavior detection algorithm, we use the collected datasets to measure its effectiveness with regard to different parameters, compare the performance with other methods, and test the power computation of our mobile app.
Specifically, three Long Short-Term Memory (LSTM) \cite{hochreiter1997long} based methods are also implemented to compare the detection accuracy.
In addition, we present the real-world deployment of our system.

\subsection{Performance Evaluation}
\textbf{Effectiveness evaluation.}
In this experiment, we evaluate the effectiveness of the intelligent behavior detection module on the whole taxi dataset and canteen dataset, respectively.
We evaluate the effectiveness from two aspects, i.e., detection accuracy (ACC) and false positive rate (FPR).
The ACC is the proportion of correct inferences for both check-out events and non-check-out events among the total number of records in the dataset.
The FPR is calculated as the ratio between the number of check-out events wrongly detected in the in-venue windows (false positives) and the total number of records in the dataset.

Three algorithmic parameters, including window width $l$, \#~continuous windows $\mathbf{L_w}$, and powered average threshold $\mathbf{L_a}$, are evaluated on their influences on the detection performance.
In the experiment, we first use grid searches to find the optimal values of these three parameters as their initial settings.
Then, we measure the effectiveness by varying the setting of each parameter separately. Additionally, we compare our proposed joint detection method with only volatility detection (VD) and only cyclicity detection (CD).

\figref{fig:parameters} (a) shows the ACC and FPR results of the three methods with regard to window width $l$ under the taxi scenario. In general, under various settings of $l$, our method can always have a higher ACC and a lower FPR, demonstrating the effectiveness of combining the volatility detection and cyclicity detection.
If $l$ is set at $2s$ (the optimal value), our joint detection method can achieve the best performance with 85.26\% ACC and 9.62\% FPR.
As a contrast, both the other two methods have worse performance with 78.21\% ACC \& 16.67\% FPR and 15.38\% ACC \& 0.00\% FPR, respectively.
From the results, we can also find the accuracy of the joint method will be decreased with a longer window width.
This is mainly because the volatility feature becomes indiscriminative with a longer period in the taxi scenario.

\figref{fig:parameters} (b) plots the ACC and FPR results with regard to \# continuous windows $\mathbf{L_w}$.
As can be seen, the best performance is achieved when three continuous check-out windows are used to detect  check-out events.
Due to the inherent volatility of taxi riding, more false positive detection cases emerge with fewer windows, resulting in a lower ACC and a higher FPR.
Moreover, benefited from the cyclicity detection, more efficient detection could be achieved with $2s \times 3 ~windows = 6s$, as compared to the case of only using volatility detection with $4s \times 3~windows = 12s$.

\figref{fig:parameters} (c) shows the ACC and FPR results with regard to the powered average threshold $\mathbf{L_a}$ used in the volatility detection.
When the threshold is set at 120, the highest accuracy and a relatively low FPR can be achieved.
The experiment results also show that our joint detection method could perform better than the volatility detection through jointly considering the cyclicity of human walking patterns.

Next, we evaluate the performance for the canteen scenario.
The experiment results are shown in \figref{fig:parameters} (d), (e), and (f).
The optimal values for window width $l$, \# continuous windows $\mathbf{L_w}$, and powered average threshold $\mathbf{L_a}$ are $4s$, 6, and 20, respectively.
With these parameters, our method can achieve the best performance (i.e., 81.82\% ACC and 6.36\% FPR) on the whole canteen dataset.
In contrast, the optimal volatility detection only achieves 77.27\% ACC and 4.55\% FPR with $8s$ window width and eight continuous windows.
These results suggest that in the canteen scenario, check-out events could be detected more efficiently with our joint method.
This is because users' movement in canteens tends to be gentle and the behavior pattern change during the check-out is distinct for detection. Thus,
benefiting from the cyclicity detection, the detection time of our method can be reduced compared with the volatility detection.
In summary, the experiment results in both scenarios show that our joint method can always achieve higher effectiveness compared with the other non-joint methods.

\begin{table*}[t]
\caption{Overall performance comparison between different methods on both taxi and canteen dataset.}
\centering
\begin{tabular}{cccccc}
\hline
 & \multicolumn{1}{c|}{} & \multicolumn{1}{c}{\multirow{2}{*}{Status Change}} & \multicolumn{1}{c}{\multirow{2}{*}{Current Status}} & Current Status & \multicolumn{1}{c}{\multirow{2}{*}{Ours}} \\
 & \multicolumn{1}{c|}{} &  &  & (Balanced) &
\\ \hline
\multicolumn{1}{c|}{\multirow{3}{*}{Taxi}} & \multicolumn{1}{l|}{Classification Accuracy} & Overfitting & Overfitting & 57.17\% & - \\
\multicolumn{1}{c|}{} & \multicolumn{1}{l|}{Classification Loss} & 0.0609 & 0.1559 & 0.1623 & - \\
\multicolumn{1}{c|}{} & \multicolumn{1}{l|}{Check-out Accuracy} & 0.00\% & 0.00\% & 40.61\% & \textbf{90.91\%}
\\ \hline
\multicolumn{1}{c|}{\multirow{3}{*}{Canteen}} & \multicolumn{1}{l|}{Classification Accuracy} & Overfitting & Overfitting & 87.99\% & - \\
\multicolumn{1}{c|}{} & \multicolumn{1}{l|}{Classification Loss} & 0.0364 & 0.0579 & 0.0722 & - \\
\multicolumn{1}{c|}{} & \multicolumn{1}{l|}{Check-out Accuracy} & 0.00\% & 0.00\% & 71.43\% & \textbf{81.14\%}
\\ \hline
\end{tabular}
\label{tab:performanceComparison}
\vspace{-1em}
\end{table*}

\textbf{Comparison with LSTM.}
To further evaluate our proposed algorithm, we compare the detection performance with three methods based on LSTM~\cite{hochreiter1997long}. For the LSTM based methods, we randomly extract two thirds of the whole dataset as the training set, and the rest as the testing set.
After training the parameters on the training set, we measure the algorithm performance on the testing set with the generated optimal parameter values.
Each accelerometer data record is divided by a sliding window with $5s$ width and further fed into the model.
Three methods are tested, including detecting the status change window (SC), detecting the current status of the window (CS), and detecting the current status of the window with balanced data (CSB).
In the SC method, windows are categorized into two classes, i.e., the status change window and the status unchange window.
The CS method aims at recognizing two types of windows in terms of the current status, including the in-venue window and the out-venue window.
For the CSB method, considering the unbalanced sample quantity, we manually synthesize data records by concatenating more out-venue windows and then classify windows as the CS method.

As shown in \tabref{tab:performanceComparison}, our proposed approach outperforms the LSTM-based methods in both scenarios.
The joint detection method achieves a higher check-out event detection accuracy on the testing dataset, because many discriminative features are successfully extracted and recognized.
In contrast, the LSTM-based methods cannot learn effective features from the unbalanced dataset or the size-limited dataset, resulting in overfitting or lower accuracy.
Specifically, for the SC and CS methods, the quantity of status change windows and out-venue windows in the dataset are both very limited.
In our experiments, the overfitting is observed, where the two methods cannot converge and only output the dominant class (i.e., the status unchange window and the in-venue window, respectively).
Thus, for the two methods, the check-out accuracy is only 0.00\% for both of our datasets (without synthesized data records).

\begin{figure}[t]
    \centering
    \includegraphics[width=0.45\textwidth]{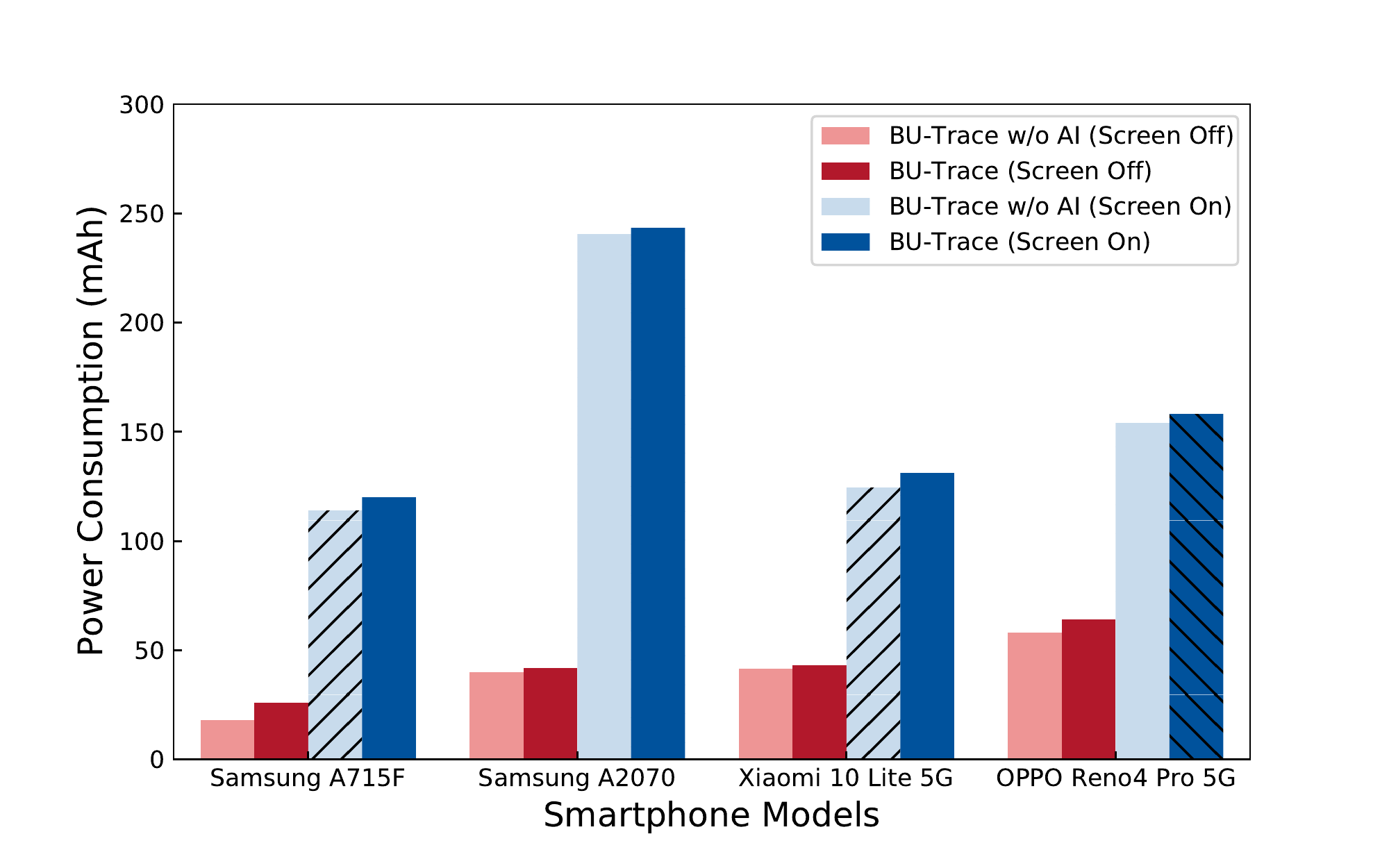}
    \caption{Power consumption of $\mathsf{BU}$-$\mathsf{Trace}$}
    \label{fig:powerConsumption}
    \vspace{-1em}
\end{figure}

\textbf{Power consumption.}
In this experiment, we test the power consumption of our system on four models of mobile phones.
Four testing cases are considered, i.e., $\mathsf{(i)}$ $\mathsf{BU}$-$\mathsf{Trace}$ without intelligent detection (screen off), $\mathsf{(ii)}$ complete $\mathsf{BU}$-$\mathsf{Trace}$ (screen off), $\mathsf{(iii)}$ $\mathsf{BU}$-$\mathsf{Trace}$ without intelligent detection (screen on), and $\mathsf{(iv)}$ complete $\mathsf{BU}$-$\mathsf{Trace}$ (screen on).
For each case, we conduct the test for one hour to obtain more stable results.
\figref{fig:powerConsumption} shows the results.
When the phones are in screen-off status, the intelligent auto check-out module consumes 4.32$mAh$ of extra power in one hour on average.
In other words, the average extra power consumed by the intelligence module is 10.96\% of the consumed power by working phones without this module.
When the phones are in screen-on status, the power consumed by screen display dominates.
The average power consumed by the intelligence module is 3.08\% of that by working phones without this module. All these results indicate that our $\mathsf{BU}$-$\mathsf{Trace}$ system does not consume  much power during user behaviour monitoring.

\begin{figure}[t]
    \centerline{
        \subfigure[$\mathsf{BU}$-$\mathsf{Trace}$]{
            \includegraphics[width=0.19\columnwidth]{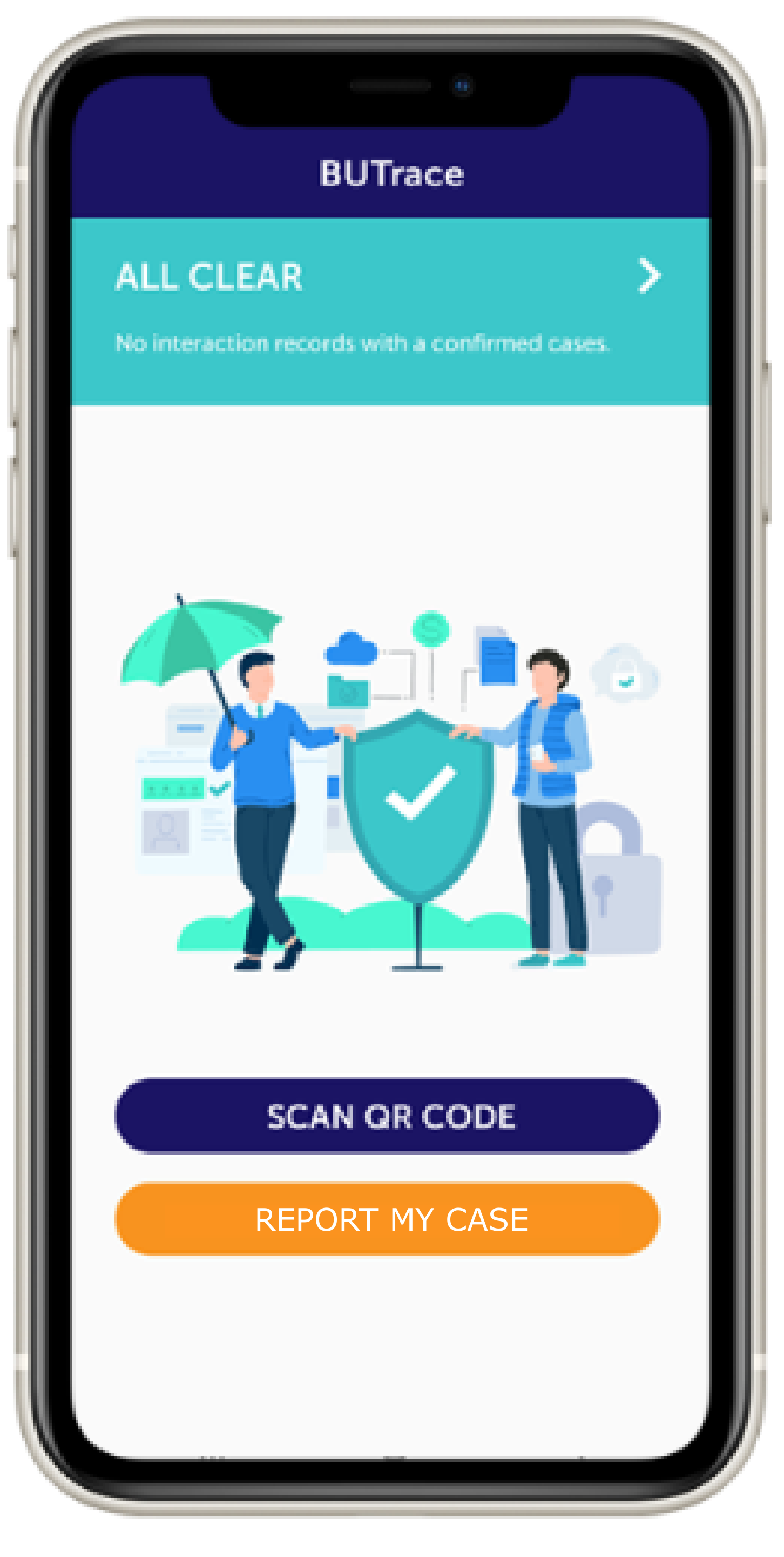}}
    \hfil
        \subfigure[Tailored QR code and NFC tag]{
            \includegraphics[width=0.76\columnwidth]{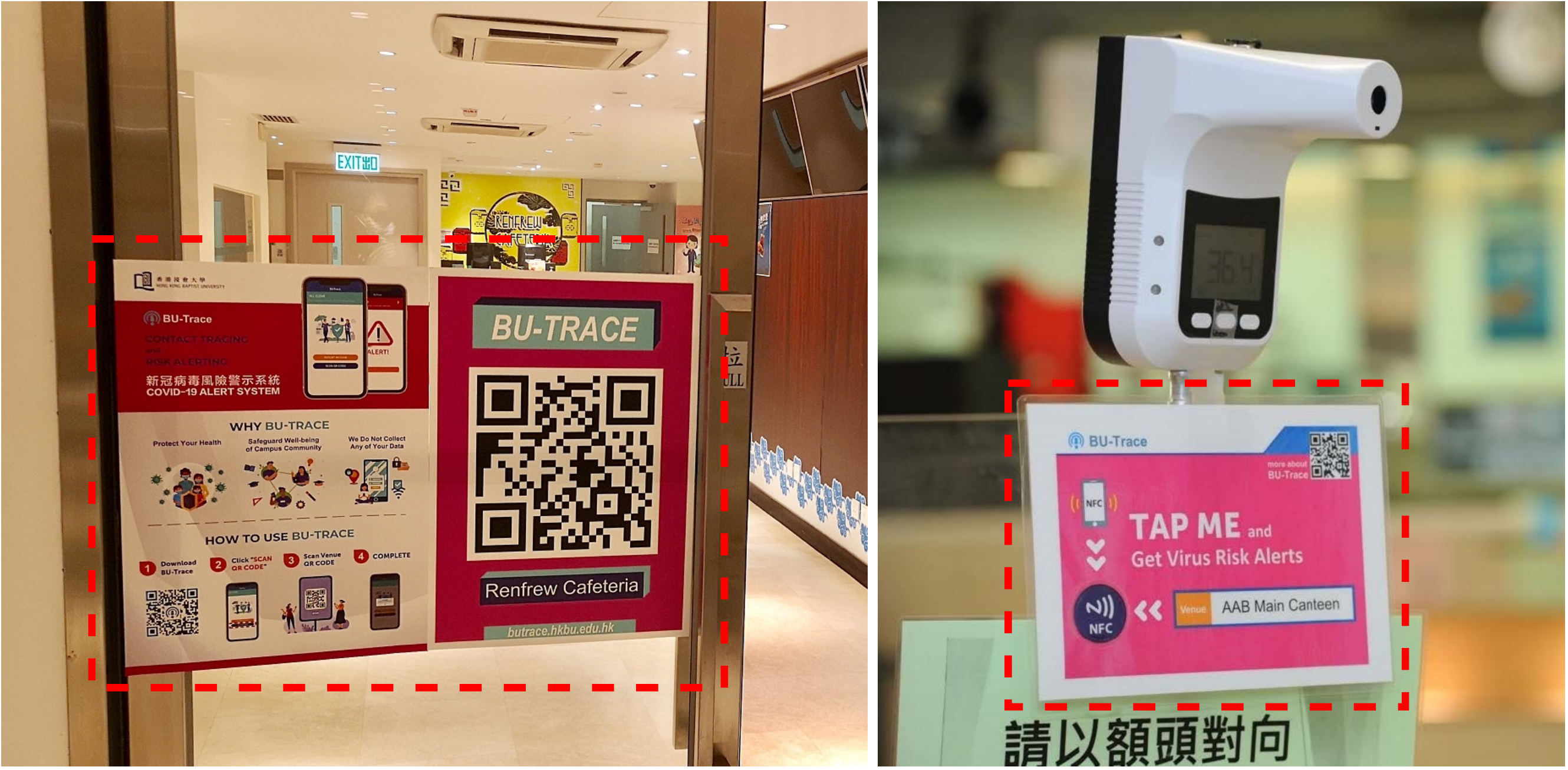}}}
    \caption{$\mathsf{BU}$-$\mathsf{Trace}$ deployment}
    \label{fig:deployment}
    \vspace{-1em}
\end{figure}

\textbf{System deployment.}
Our developed $\mathsf{BU}$-$\mathsf{Trace}$ system has been deployed on the university campus since September 2020~\cite{BUTrace}.
The mobile app is available for download from both Google Play (for Android users) and Apple's App Store (for iOS users).
Students and staff can install and use the mobile app without enabling the location permission.
\figref{fig:deployment}(a) shows the main interface of the $\mathsf{BU}$-$\mathsf{Trace}$ app.
\figref{fig:deployment}(b) shows some photos of the tailored QR code and NFC tag at a venue entrance.
To effectively realize contact tracing, over 200 QR codes and NFC tags have been deployed at selected venues, such as canteens, libraries, labs, and meeting rooms. The system has been successful in tracing the contacts of an infected student and protecting the safety of the campus community.

\section{Conclusion}
In this paper, we propose $\mathsf{BU}$-$\mathsf{Trace}$, a permissionless mobile system for contact tracing based on QR code and NFC technologies.
Compared with previous works, $\mathsf{BU}$-$\mathsf{Trace}$ offers user privacy protection and system intelligence without requesting location or other privacy-related permissions.
To realize this system, a user study is first conducted to investigate and quantify the user acceptance of a mobile contact tracing system.
Then, a decentralized system is proposed to enable contact tracing while protecting user privacy.
Finally, an intelligent behavior detection algorithm is developed to ease the use of our system.
We implement $\mathsf{BU}$-$\mathsf{Trace}$ and conduct extensive experiments in practical scenarios.
The evaluation results demonstrate the effectiveness of our proposed permissionless mobile system.

\section*{Acknowledgement}

This research is supported by a strategic development grant from Hong Kong Baptist University.

\addtolength{\textheight}{-14.9cm} 

\bibliographystyle{ieeetr}
\bibliography{exbible}

\begin{thebibliography}{10}

\bibitem{flaxman2020estimating}
S.~Flaxman, S.~Mishra, A.~Gandy, H.~J.~T. Unwin, {\em et~al.}, ``Estimating the
  effects of non-pharmaceutical interventions on covid-19 in europe,'' {\em
  Nature}, vol.~584, no.~7820, pp.~257--261, 2020.

\bibitem{ferretti2020quantifying}
L.~Ferretti, C.~Wymant, M.~Kendall, L.~Zhao, {\em et~al.}, ``Quantifying
  sars-cov-2 transmission suggests epidemic control with digital contact
  tracing,'' {\em Science}, vol.~368, no.~6491, 2020.

\bibitem{zeinalipour2020covid}
D.~Zeinalipour-Yazti and C.~Claramunt, ``Covid-19 mobile contact tracing apps
  (mcta): A digital vaccine or a privacy demolition?,'' in {\em Proc. of IEEE
  MDM}, 2020.

\bibitem{li2012nearby}
H.~P. Li, H.~Hu, and J.~Xu, ``Nearby friend alert: Location anonymity in mobile
  geosocial networks,'' {\em IEEE Pervasive Computing}, vol.~12, no.~4,
  pp.~62--70, 2012.

\bibitem{healthCode}
P.~Mozur {\em et~al.}, ``In coronavirus fight, china gives citizens a color
  code, with red flags,'' 2020.

\bibitem{koreanApp}
H.~Jeremy, ``Contact tracing apps struggle to be both effective and private,''
  {\em IEEE Spectrum}, 2020.

\bibitem{indiaApp}
``{Aarogya Setu}.'' \url{https://www.mygov.in/aarogya-setu-app/}.

\bibitem{safeEntry}
``{SafeEntry}.'' \url{https://www.safeentry.gov.sg}.

\bibitem{bay2020bluetrace}
J.~Bay, J.~Kek, A.~Tan, C.~S. Hau, L.~Yongquan, {\em et~al.}, ``{BlueTrace}: A
  privacy-preserving protocol for community-driven contact tracing across
  borders,'' {\em Government Technology Agency-Singapore, Tech. Rep}, 2020.

\bibitem{covidSafe}
``{COVIDSafe}.''
  \url{https://www.health.gov.au/resources/apps-and-tools/covidsafe-app}.

\bibitem{canadaApp}
``{COVID Shield}.''
  \url{https://www.nhsinform.scot/illnesses-and-conditions/infections-and-poisoning/coronavirus-covid-19/coronavirus-covid-19-shielding}.

\bibitem{germanApp}
``{Corona-Warn}.''
  \url{https://www.bundesregierung.de/breg-de/themen/corona-warn-app/corona-warn-app-englisch}.

\bibitem{swissApp}
``{SwissCovid}.''
  \url{https://www.bag.admin.ch/bag/en/home/krankheiten/ausbrueche-epidemien-pandemien/aktuelle-ausbrueche-epidemien/novel-cov/swisscovid-app-und-contact-tracing.html}.

\bibitem{ens}
``{Exposure Notifications}.''
  \url{https://www.google.com/covid19/exposurenotifications/}.

\bibitem{cna2020news}
``{TraceTogether}.''
  \url{https://www.channelnewsasia.com/news/singapore/covid-19-singapore-low-community-prevalence-testing-13083194}.

\bibitem{peng2018indoor}
Z.~Peng, S.~Gao, B.~Xiao, G.~Wei, S.~Guo, and Y.~Yang, ``Indoor floor plan
  construction through sensing data collected from smartphones,'' {\em IEEE
  IoTJ}, vol.~5, no.~6, pp.~4351--4364, 2018.

\bibitem{gonzalez2017direct}
J.~A. Gonzalez, L.~A. Cheah, {\em et~al.}, ``Direct speech reconstruction from
  articulatory sensor data by machine learning,'' {\em IEEE/ACM Trans. on
  ASLP}, vol.~25, no.~12, pp.~2362--2374, 2017.

\bibitem{yao2017efficient}
Y.~Yao {\em et~al.}, ``An efficient learning-based approach to multi-objective
  route planning in a smart city,'' in {\em Proc. of IEEE ICC}, 2017.

\bibitem{shi2018machine}
X.~Shi and D.-Y. Yeung, ``Machine learning for spatiotemporal sequence
  forecasting: A survey,'' {\em arXiv preprint arXiv:1808.06865}, 2018.

\bibitem{medsker2001recurrent}
L.~R. Medsker and L.~Jain, ``Recurrent neural networks,'' {\em Design and
  Applications}, vol.~5, 2001.

\bibitem{hochreiter1997long}
S.~Hochreiter {\em et~al.}, ``Long short-term memory,'' {\em Neural
  computation}, vol.~9, no.~8, pp.~1735--1780, 1997.

\bibitem{santos2019artificial}
O.~C. Santos, ``Artificial intelligence in psychomotor learning: modeling human
  motion from inertial sensor data,'' {\em IJAIT}, vol.~28, no.~04, p.~1940006,
  2019.

\bibitem{shoaib2014fusion}
M.~Shoaib, S.~Bosch, O.~D. Incel, H.~Scholten, and P.~J. Havinga, ``Fusion of
  smartphone motion sensors for physical activity recognition,'' {\em Sensors},
  vol.~14, no.~6, pp.~10146--10176, 2014.

\bibitem{bedogni2012train}
L.~Bedogni, M.~Di~Felice, and L.~Bononi, ``By train or by car? detecting the
  user's motion type through smartphone sensors data,'' in {\em IEEE Wireless
  Days}, 2012.

\bibitem{fang2016transportation}
S.-H. Fang, H.-H. Liao, Y.-X. Fei, K.-H. Chen, J.-W. Huang, Y.-D. Lu, and
  Y.~Tsao, ``Transportation modes classification using sensors on
  smartphones,'' {\em Sensors}, vol.~16, no.~8, p.~1324, 2016.

\bibitem{rein2010low}
S.~Rein and M.~Reisslein, ``Low-memory wavelet transforms for wireless sensor
  networks: A tutorial,'' {\em IEEE Communications Surveys \& Tutorials},
  vol.~13, no.~2, pp.~291--307, 2010.

\bibitem{BUTrace}
``{BU-Trace}.'' \url{https://butrace.hkbu.edu.hk/}.

\end{thebibliography}

\end{document}